# Optical Klystron Enhancement to SASE X-ray FELs

Yuantao Ding,[*] Paul Emma, and Zhirong Huang[+]

Stanford Linear Accelerator Center, Menlo Park, California 94025, USA

Vinit Kumar [a]

Argonne National Laboratory, Argonne, Illinois 60439, USA

The optical klystron enhancement to self-amplified spontaneous emission (SASE) free electron lasers (FELs) is studied in theory and in simulations. In contrast to a seeded FEL, the optical klystron gain in a SASE FEL is not sensitive to any phase mismatch between the radiation and the microbunched electron beam. The FEL performance with the addition of four optical klystrons located at the undulator long breaks in the Linac Coherent Light Source (LCLS) shows significant improvement if the uncorrelated energy spread at the undulator entrance can be controlled to a very small level. In addition, FEL saturation at shorter x-ray wavelengths (around 1.0 Å) within the LCLS undulator length becomes possible. We also discuss the application of the optical klystron in a compact x-ray FEL design that employs relatively low electron beam energy together with a short-period undulator.

## I Introduction

An x-ray free electron laser (FEL) operated in the self-amplified spontaneous emission (SASE) mode is the primary candidate for the next-generation light source and is under active development around the world [1,2,3]. In such a device, a high-brightness electron beam passing a long undulator develops energy and density modulations at the radiation wavelength and consequently amplifies the spontaneous emission into intense, coherent

---

[*] Electronic address: ding@slac.stanford.edu

[+] Electronic address: zrh@slac.stanford.edu

[a] Current address: G-20, ADL Building, Raja Ramanna Centre for Advanced Technology, Indore-452013, INDIA.

radiation. Based on the achievable electron beam qualities such as peak current and transverse emittances, the total length of the undulator required to reach the x-ray intensity saturation usually exceeds 100 m. The electron beam energy spread is typically too small to affect the SASE performance.

To enhance the FEL gain, the optical klystron concept has been invented by Vinokurov and Skrinsky [4] and has been successfully implemented in many FEL oscillator facilities such as the Duke FEL [5]. An optical klystron consists of two undulators, separated by a dispersive section (a magnetic chicane). The dispersive section converts beam energy modulation into density modulation and hence speeds up the gain process. Theoretical studies of the optical klystron in high gain FEL amplifiers [6,7,8] show that its performance depends critically on the electron beam energy spread. More recently, Neil and Freund [9] have studied a distributed optical klystron configuration using the LCLS parameters. Based on the FEL amplifier simulations that start with a coherent seed, they point out that the performance of the optical klystron for short-wavelength FELs is very sensitive to the exact slippage of the electron beam relative to the radiation in the dispersive section. Thus, the magnetic fields of the chicane must be carefully designed and controlled to very high precision.

Motivated by the very small uncorrelated energy spread of the electron beam that has been measured in a photocathode rf gun [10], we study the possible optical klystron enhancement to SASE x-ray FELs. In Sec. II, we generalize the previous high-gain optical klystron theory to a SASE FEL having a wide bandwidth. We show that a SASE optical klystron is not sensitive to the relative phase of the electron beam to the radiation as long as the electron slippage length in the dispersive section is much longer than the coherence length of the radiation. In Sec. III, we use the LCLS as a typical x-ray FEL and discuss the evolution and the control of the uncorrelated energy spread in the accelerator and the undulator. Based on extensive SASE simulations, we illustrate the gain enhancement of the optical klystron to the LCLS and apply this method to extend its x-ray wavelength reach. We also discuss the application of the optical klystron in a compact x-ray FEL design that employs relatively low beam energy together with a shorter-period undulator. Finally, we summarize our studies and conclude that the optical klystron is a promising approach to enhance the x-ray FEL performance.

## II. One-dimensional analysis

In this section, we analyze an optical klystron configuration with a magnetic chicane between two high-gain FEL undulators and extend the previous theoretical treatments [6-8] to the SASE operating mode. Saldin et al. recently consider an FEL klystron amplifier that uses an uncompressed electron bunch with a relatively low current and an extremely small energy spread [11]. Thus, the first undulator in their proposal is a pure energy modulator (no gain in radiation), while the density modulation is generated only after the beam passes the chicane. Although their proposal is conceptually simple, the relatively low current of the electron beam is not capable of driving a hard x-ray FEL. Thus, in this and the following sections, we focus our attention on the study of optical klystrons in high-gain x-ray FELs.

A magnetic chicane introduces an energy-dependent longitudinal delay of the electron relative to the radiation, which can be expressed as a change of the radiation phase "seen" by the electron:

$$\Delta \theta = -\frac{k_r R_{56}}{2} + k_r R_{56} \delta .\qquad(1)$$

Here $\lambda_r = 2\pi / k_r = 2\pi c / \omega_r$ is the FEL resonant wavelength, $R_{56}$ is the momentum compaction of the chicane, and $\delta = (\gamma - \gamma_0)/\gamma_0$ is the relative energy deviation. The first term in Eq. (1) describes the overall phase slippage between the FEL radiation and the reference electron having the energy $\gamma_0 mc^2$, and the second term describes the relative phase change for an electron with a slightly different energy. Following the one-dimensional (1D) theory of Kim [8] but keep the overall phase slippage, we write down the optical klystron (OK) enhancement factor to the radiation field $E_\nu$ at the scaled frequency $\nu = \omega / \omega_r$:

$$R(\nu) \equiv \frac{E_\nu^{OK}}{E_\nu^{no\ OK}} = \frac{1 - \int d\xi \frac{dV(\xi)/d\xi}{(\mu-\xi)^2} e^{-i\rho k_r \nu R_{56} \xi} e^{i k_r \nu R_{56}/2}}{1 + 2 \int d\xi \frac{V(\xi)}{(\mu-\xi)^3}} ,\qquad(2)$$

where $\xi = \delta / \rho$ is the normalized energy variable, $\rho$ is the FEL Pierce parameter [12], $\mu$ is the complex growth rate of the radiation field in each undulator ($E_\nu \propto \exp(-i 4\pi \mu \rho N_u)$, with $N_u$ being the number of undulator periods), $\mu = (-1 + i\sqrt{3})/2$ for a beam with a vanishing energy spread, and $V(\xi)$ is the energy distribution of the electron beam with the normalization $\int V(\xi) d\xi = 1$.

The first term in the numerator of Eq. (2) represents the contribution from the radiation in the first undulator, while the second term in the numerator represents the contribution of the microbunched electron beam. The last exponent of the second term (i.e., $\exp(i k_r \nu R_{56}/2)$) represents the "on-energy" phase slippage of the electron beam relative to the radiation due to the chicane. For extremely small beam energy spread the second term (microbunching) dominates over the first term (radiation), and there is no need for phase matching between the two terms. For the optical klystron FELs considered in this paper, the energy spread practically limits the amount of the microbunching induced by the chicane, and hence both terms in Eq. (2) must be taken into account. For a seeded FEL with $v = 1$, $k_r R_{56}/2 = 2\pi n$ ($n = 1,2,3,\ldots$), this yields a nearly matched phase (i.e., constructive interference between two terms). The optical klystron is then optimized as assumed in Refs. [6-8]. However, in the hard x-ray wavelength range, changing $R_{56}$ of the chicane by a fraction of 1 Angstrom can result in a complete phase mismatch. Thus, there can be large fluctuations in the radiation power due to small fluctuations in the magnetic fields as observed in Ref. [9], especially when more than one optical klystron is used in a distributed optical klystron configuration.

Nevertheless, SASE FELs start from shot noise and have a relatively wide bandwidth. For a given value of $R_{56}$, the phase may be mismatched for one particular wavelength but may be properly matched for another wavelength within the SASE bandwidth. Thus, we should integrate over the SASE spectrum $S(v)$ to obtain the optical klystron power gain factor as:

$$G = \int dv |R(v)|^2 S(v). \tag{3}$$

Here we can use the average SASE spectrum given by

$$S(v) = \frac{1}{\sqrt{2\pi}} \exp\left[-\frac{(v-1)^2}{2\sigma_v^2}\right], \tag{4}$$

where $\sigma_v$ is the relative rms bandwidth that decreases with increasing undulator distance up to the saturation.

For an electron beam with a Gaussian energy distribution of rms width $\sigma_\delta \ll \rho$ (i.e., $\sigma_\xi \ll 1$), we can integrate Eq. (2) over energy and Eq. (3) over frequency to obtain

$$G \approx \frac{1}{9}\left[\begin{array}{l} 5 + D^2 \exp(-D^2\sigma_\xi^2) + 2\sqrt{3}D\exp\left(-\frac{D^2\sigma_\xi^2}{2}\right) \\ + \left(4 + \sqrt{3}D\exp\left(-\frac{D^2\sigma_\xi^2}{2}\right)\cos\left(\frac{D}{2\rho}\right) - D\exp\left(-\frac{D^2\sigma_\xi^2}{2}\right)\sin\left(\frac{D}{2\rho}\right)\right)\exp\left(-\frac{D^2\sigma_v^2}{8\rho^2}\right) \end{array}\right], \tag{5}$$

where $D = k_r R_{56}\rho$. The power gain factor $G$ as a function of the chicane strength $R_{56}$ is shown in Fig. 1 for two typical values of the rms energy spread $\sigma_\delta$, assuming a typical rms SASE bandwidth $\sigma_v = \rho$. When $R_{56}$ is very small, the optical klystron operates as a phase shifter, and the FEL power is oscillatory depending on the relative phase between the radiation and the electron beam. As $k_r R_{56}\sigma_\delta \to 1$, the optical klystron gain peaks and starts to decay exponentially due to the smearing effect of the intrinsic energy spread. In addition, the power oscillation damps with increasing $R_{56}$. Thus, the phase matching is no longer important when the optical klystron is near its peak performance. As the optimal $R_{56}$ is given by $k_r R_{56} \sigma_\delta \sim 1$, the damping of the last (oscillatory) term in Eq.(5) is effective when $D = k_r R_{56} \rho \gg 1$. The last inequality is always satisfied since $\sigma_\delta \ll \rho$ is a necessary condition for the application of a high-gain optical klystron [6]. Thus, we can always ignore the last term in Eq. (5) and arrive at a simplified gain formula as

$$G \approx \frac{1}{9}\left[5 + D^2 \exp\left(-D^2 \sigma_\xi^2\right) + 2\sqrt{3} D \exp\left(-\frac{D^2 \sigma_\xi^2}{2}\right)\right]. \tag{6}$$

A simple physical picture emerges in the time domain. The path length difference between the SASE radiation and the electron beam passing the dispersive section is about $R_{56}/2 \approx 1/(2 k_r \sigma_\delta) = \lambda_r/(4\pi \sigma_\delta)$ at the optimal chicane setting. Since the typical SASE coherence length is on the order of $\lambda_r/(4\pi \rho)$ [13,14], it is much smaller than the path length difference introduced by the chicane for the beam with $\sigma_\delta \ll \rho$. Therefore, there is no place for the electron beam to match the radiation phase after the beam is slipped from the SASE radiation more than a few temporal spikes. The radiation power averaged over many statistically independent spikes is then not sensitive to the exact slippage introduced by the chicane. This feature distinguishes an optical klystron in a SASE FEL from that in a seeded FEL, which is always subject to phase matching unless the dispersively enhanced microbunching dominates over the radiation by more than an order of magnitude.

## III. Three-dimensional simulations

In this section, we first study the evolution and the control of the uncorrelated energy spread in the accelerator and the undulator, which plays a crucial role in determining the optical klystron performance. We then use three-dimensional (3D) simulations to explore the LCLS gain enhancement with a distributed optical klystron configuration for two different radiation wavelengths of 1.5 Å and 1.0 Å. The phase of one optical klystron is independently varied in the simulations in order to verify that the output power is not sensitive to any phase mismatch, as predicted in the above 1D analysis. Finally, we discuss the optical klystron enhancement to a compact x-ray FEL using a relatively low-energy beam together with a short-period undulator.

**A. Uncorrelated energy spread of the LCLS beam**

As discussed in the 1D analysis, the uncorrelated energy spread plays a crucial role for the gain enhancement of the optical klystron. To satisfy the condition $\sigma_\delta \ll \rho$, we analyze the smallest possible energy spread for the LCLS electron beam. Two main sources of energy spread are considered, one is from the gun and the linac, which forms the initial energy spread at the entrance of the FEL undulator; while the other is the quantum diffusion due to spontaneous radiation along the undulator, which leads to an increase of energy spread after the electron beam is injected into the undulator. Since the proposed optical klystrons operate in the early stage of the exponential regime, the FEL-induced energy spread is negligible (but is included in the simulations).

The uncorrelated energy spread of electron beams generated from a photocathode rf gun can be extremely small, at an rms value of 3 to 4 keV from both measurements [10] and analysis[15]. Nevertheless, a microbunching instability driven by coherent synchrotron radiation [16,17,18,19] and longitudinal space charge [20,21] in the accelerator system may be large enough to significantly degrade the beam qualities including the energy spread. This microbunching instability occurs at much longer wavelengths than the FEL microbunching and requires much larger $R_{56}$ (from bunch compressor chicanes) than the optical klystron chicanes. To maintain a relatively small energy spread after compression and acceleration, both a smooth drive-laser profile and a low microbunching gain are necessary. Huang et al [20] discussed in detail the suppression of the microbunching instability in the LCLS, where a laser heater [19] was adapted to provide strong Landau damping against the instability without degrading the SASE performance. For the LCLS at 1.5 Å, the tolerable rms energy spread at the undulator entrance is $1\times10^{-4}$ at 14 GeV, which is about $0.2\rho$ as $\rho \approx 5\times10^{-4}$. By using the laser heater to increase the rms energy spread from 3 to 40 keV in the LCLS injector, after a total compression factor of about 30, the slice rms energy spread at the undulator entrance can be controlled to $1\times10^{-4}$ [20]. However, considering the gain enhancement of the optical klystron (see Fig. 1), a smaller energy spread (e.g., $5\times10^{-5}$ or $0.1\rho$) is desirable. This may be achievable by dropping the heater-induced energy spread to 20 keV at the expense of the increased microbunching instability gain. Figure 2 shows the expected microbunching gain with respect to a small initial density modulation at the injector end for the 1-nC nominal LCLS bunch at these energy spread levels. The instability tolerance to this smaller energy spread depends significantly on the smoothness of the drive-laser profile and may be tested experimentally in the LCLS. A smooth Gaussian drive-laser profile may be more desirable in this case than a flattop profile with many intensity ripples. Recently, a low-charge LCLS option is proposed to mitigate collective effects while maintain a similar FEL performance [22]. The microbunching instability gain under this low-charge configuration is much smaller even at the reduced energy spread controlled by the laser heater.

Another source of the energy spread, the energy diffusion due to spontaneous radiation along the undulator, was discussed by Saldin et al [23]. For a planar undulator, the energy diffusion is given by:

$$\frac{d<(\Delta\gamma)^2>}{dz} = \frac{7}{15}\frac{\lambda_c}{2\pi}r_e\gamma^4 k_u^3 K^2 F(K), \tag{7}$$

where $\lambda_c$ is the Compton wavelength, $r_e$ is the classical radius of the electron, $k_u = 2\pi/\lambda_u$ with $\lambda_u$ being the undulator period, $K = eB_0/(mck_u)$ is the undulator parameter with a peak magnetic field $B_0$, $F(K) \approx 1.2K$ for $K \gg 1$. It is important to stress that the energy diffusion rate increases with $\gamma^4$ and $K^3$. Thus, a simultaneous reduction in beam energy and undulator parameter can significantly reduce this effect. For the LCLS at $\lambda_r$ = 1.5 Å and $K$=3.5, the rms energy spread increases from initial value of $5\times10^{-5}$ to $1\times10^{-4}$ at the undulator position of 40 m due to the spontaneous radiation. We note that this effect is not included in the LCLS simulations presented in Ref. [9], but is included in all our FEL simulations to be discussed below.

**B. LCLS simulation studies**

We use GENESIS 1.3 [24] to perform the 3D simulations of the LCLS optical klystrons as well as a compact x-ray FEL (see Sec. III C). The main parameters used in simulations are listed in Table 1. According to the LCLS undulator configuration, there is a long break of about 1 meter between every third undulator section, where chicane structures can be installed without changing the present undulator placement. We place four 4-dipole chicanes in the first four long breaks between undulator sections (at 12, 24, 36, and 48 m) to form a distributed optical klystron configuration. For each chicane, the optimal gain enhancement is obtained by scanning the chicane dipole magnetic field strength. Two initial rms energy spread values of $1\times10^{-5}$ and $5\times10^{-5}$ at the entrance of the undulator are used in the 3D simulations. While we consider the energy spread of $5\times10^{-5}$ may be achievable in the LCLS with a smooth drive-laser profile or with the low-charge option, the energy spread of $1\times10^{-5}$ requires to switch off the laser heater completely and is probably not allowed by the microbunching instability in the linac. It is still included in the simulations in order to study the best possible optical klystron performance and the influence of spontaneous energy diffusion in the undulator. We also note that an initial rms energy spread of $1\times10^{-4}$ does not yield significant FEL improvement (or degradation) in our optical klystron configuration. In the case without any optical klystron, the simulation results show little difference between initial energy spread of $5\times10^{-5}$ and $1\times10^{-5}$.

**TABLE 1. Main simulation parameters for optical klystron x-ray FELs.**

| Parameter | unit | Fig.3 | Fig.4 | Fig.5 | Fig.7 | Fig.8 |
|---|---|---|---|---|---|---|
| Electron Energy | GeV | 13.6 | 13.5 | 11.0 | 13.6 | 5.0 |
| Normalized rms Emittance | μm | 1.2 | 1.2 | 1.2 | 1.2/1.5 | 1.0 |
| Peak Current | kA | 3.4 | 3.4 | 3.4 | 3.4 | 2.0 |
| Initial rms Energy Spread | | $5\times10^{-5}$ | $5\times10^{-5}$ | $5\times10^{-5}$ | $5\times10^{-5}$ | $2\times10^{-5}$ |
| Undulator Parameter | | 3.5 | 2.7 | 2.7 | 3.5 | 1.3 |
| Undulator Period | cm | 3.0 | 3.0 | 3.0 | 3.0 | 1.5 |
| FEL Wavelength | Å | 1.5 | 1.0 | 1.5 | 1.5 | 1.5 |

Figure 3 shows the FEL power gain along the undulator with and without optical klystrons at the resonant wavelength of 1.5 Å for $K = 3.5$ (the current LCLS design parameters). The saturation length is shortened by 13 m using these optical klystrons with an initial energy spread of $5\times10^{-5}$ and $R_{56}$ of the chicanes at around 0.25 μm (with a small variation for each chicane). We also note that a 10% variation of the chicane $R_{56}$ values does not make a visible difference for the FEL output power.

To allow for the LCLS to reach 1.0 Å without increasing the beam energy, the undulator gap may be increased by 2 mm to reduce the undulator parameter to $K = 2.7$. The 3D simulation results are presented in figure 4. Without any optical klystron, the nominal LCLS beam cannot reach SASE saturation at this wavelength. With the addition of four optical klystrons as described here, the saturation distance is shortened by about 26 m and is well within the LCLS total undulator length. At this $K$ value and using a lower beam energy (11.0 GeV), simulations of Fig. 5 also show the FEL saturation at 1.5 Å. In this case approximately 25 m of saturation length can be saved as compared to that without any optical klystron. It is clear from these numerical examples that a simultaneous reduction in beam energy and undulator parameter for the same radiation wavelength is beneficial for the optical klystron enhancement, where the energy diffusion due to spontaneous radiation in the undulator is much reduced.

Based on the LCLS parameters, the phase matching issue is also studied in 3D simulations in order to verify the 1D results discussed in section II. In GENESIS 1.3, the alignment of the radiation field and the electron beam can be controlled by input parameters. We choose the second optical klystron arrangement with 1.5-Å FEL ($K = 3.5$, initial rms energy spread of $1\times10^{-5}$) to study the FEL power fluctuations by introducing an additional phase shift in the simulations based on a particular $R_{56}$. Figure 6 shows the influence of the additional phase shifts on power. In the seeded mode, with the optimal $R_{56}$ of 0.3 μm, we observe large oscillations when this phase shift varies from 0 to $4\pi$ (corresponding to variation of $R_{56}$ from 0.3 μm to 0.3006 μm). For the SASE mode, there are very small fluctuations with these additional phase shifts at the optimal $R_{56}$ value of 0.3 μm. When we reduce $R_{56}$ to a smaller value of 0.1 μm, the power gain is reduced and the fluctuations get bigger. These 3D simulation results are in accordance with that from our 1D analysis.

Another advantage of the optical klystron scheme is to relax the electron beam emittance requirement. We study this case with the LCLS design wavelength of 1.5 Å, with a peak current of 3.4 kA and a normalized rms emittance of 1.2 μm. If the normalized emittance is relaxed to 1.5 μm, and four optical klystrons are used to enhance the bunching, the saturation length is almost the same as the case without any optical klystron but with a smaller normalized emittance at 1.2 μm, as shown in figure 7. We also note that the LCLS beam with a normalized emittance of 1.5 μm without any optical klystron will not produce saturation within the present undulator length. Thus, the optical klystron configuration relaxes the emittance requirement by more than 20%.

**C. A compact x-ray FEL**

We have seen from the previous discussions that lower electron energy and smaller undulator parameter are beneficial for reducing the energy diffusion from spontaneous radiation along the undulator. Inspired by the Spring-8 Compact SASE Source (SCSS) design [3], we study the possibility of using a relatively low energy electron beam together with a short-period undulator to drive a compact x-ray FEL with the aid of the distributed optical klystrons. As shown in Table 1, a 1.5-cm period in-vacuum undulator with $K = 1.3$ is used according to the design parameters in SCSS. To produce 1.5-Å FEL radiation, the necessary electron energy is about 5 GeV. Rather than a standard peak current of 3 kA as described in Ref [3], we assume a lower peak current of 2 kA and an rms energy spread of 100 keV (or $2\times10^{-5}$) at the undulator entrance. A smaller peak current allows for a smaller energy spread and may also help reduce the microbunching instability gain in the accelerator, as well as any wakefield effect in the small gap, in-vacuum undulator. Figure 8 shows the simulation results for the SASE mode without any optical klystron (for both 3-kA and 2-kA bunches) and with four optical klystrons (for a

2-kA bunch). The latter saturates at around 50 m of the undulator distance, which is still about 10 m shorter than the higher-current case without any optical klystron.

## VI Summary

The small, experimentally measured uncorrelated energy spread from rf guns offers the opportunity to consider applications of optical klystrons in x-ray FELs. In contrast to a seeded FEL, our paper shows that the optical klystron gain is not sensitive to the relative phase between the SASE radiation and the electron beam, and that the radiation power is very stable with a relatively large tuning range of optical klystrons. 3D simulations of the LCLS with a distributed optical klystron configuration show significant gain enhancement if the slice energy spread at the undulator entrance can be controlled to a very small level. The improved performance can be used to obtain the FEL saturation at shorter x-ray wavelengths for a fixed undulator length or to relax the stringent requirement on the beam emittance. The exploration of optical klystrons in a very compact x-ray FEL also indicates promising results. Therefore, we think that the optical klystron configuration can be an easy "add-on" to SASE x-ray FELs provided that electron beams with very small energy spreads are obtainable at the final beam energy.

**ACKNOWLEDGMENTS**

We thank J. Wu and S. Reiche for useful discussions, J. Galayda and R. Ruth for their encouragement on this work. This work was supported by Department of Energy Contracts No. DE-AC02-76SF00515.

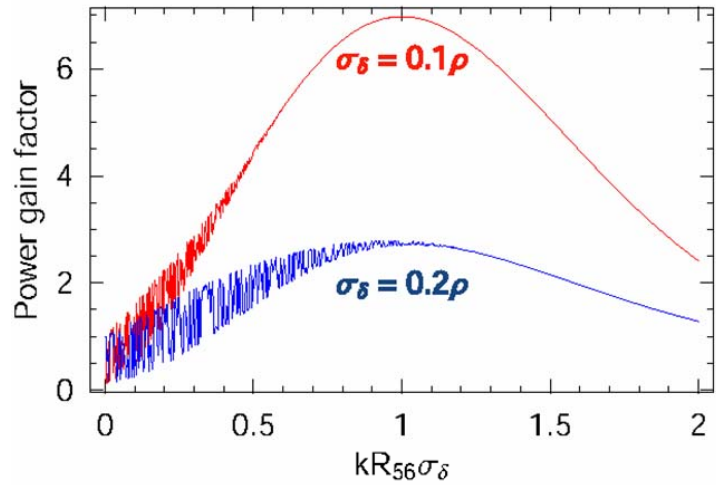

Fig. 1.(color) 1D power gain factor with relative energy spread $\sigma_\delta = 0.1\rho$ (red line) and $\sigma_\delta = 0.2\rho$ (blue line).

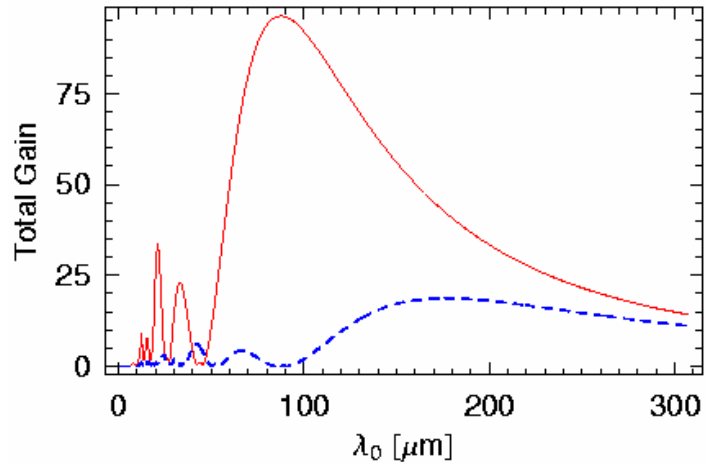

Fig. 2. (color) Total LCLS microbunching gain after two bunch compressors as a function of the initial modulation wavelength $\lambda_0$ for the laser-heated-induced rms energy spread of 20 keV (solid red curve) and rms energy spread of 40 keV (blue dashed line).The bunch charge is 1 nC.

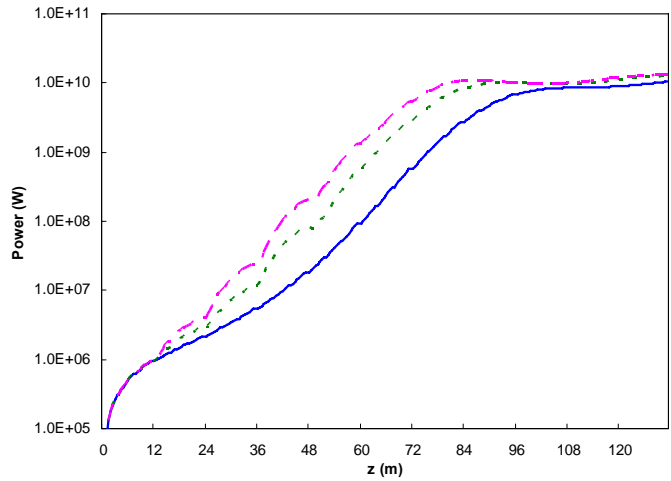

Fig. 3. (color) SASE FEL power along the undulator without any optical klystron (blue solid curve), and with 4 optical klystrons for the initial rms energy spread of $1\times10^{-5}$ (pink dashed curve) and $5\times10^{-5}$ (green dotted curve). The FEL wavelength is 1.5Å, and the undulator parameter K=3.5.

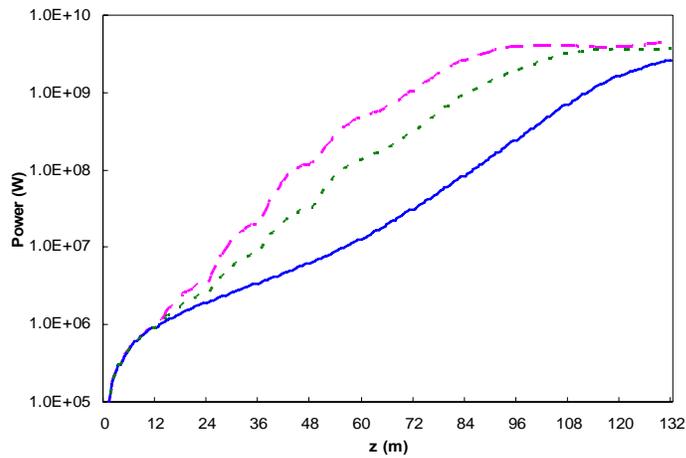

Fig. 4. (color) SASE FEL power along the undulator without any optical klystron (blue solid curve), and with 4 optical klystrons for the initial rms energy spread of $1\times10^{-5}$ (pink dashed curve) and $5\times10^{-5}$ (green dotted curve). The FEL wavelength is 1.0Å, and the undulator parameter K=2.7.

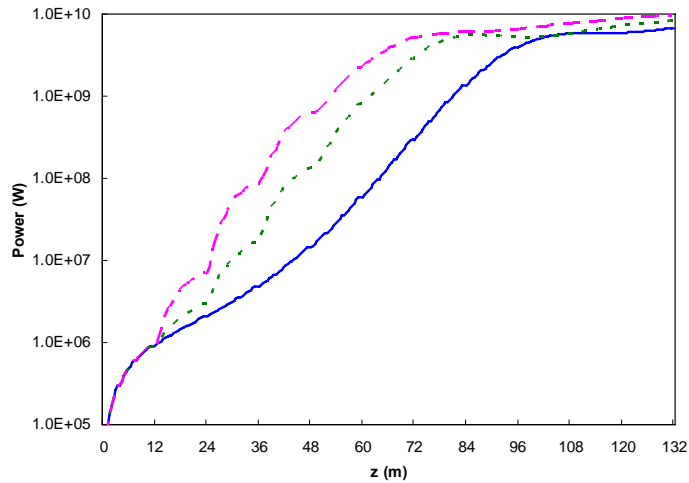

Fig. 5. (color) SASE FEL power gain along the undulator without any optical klystron (blue solid curve), and with 4 optical klystrons for the initial rms energy spread of $1\times10^{-5}$ (pink dashed curve) and $5\times10^{-5}$ (green dotted curve). The FEL wavelength is 1.5Å, and the undulator parameter K=2.7.

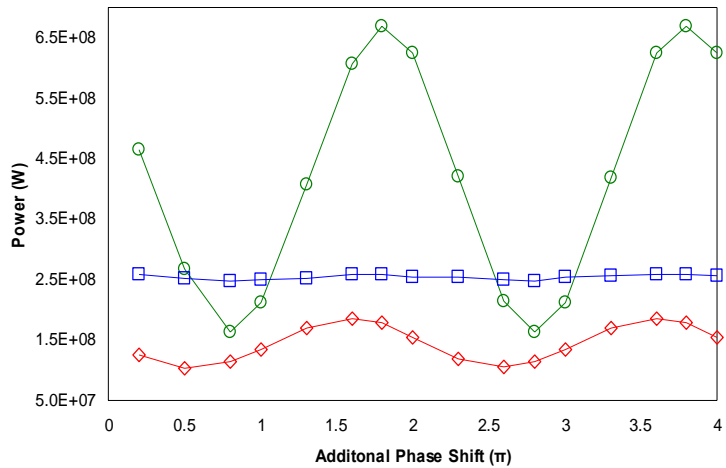

Fig 6. (color) Power fluctuations at the exponential region as a function of the additional phase shift (see text for details) based on an optimal $R_{56}$ = 0.3 μm in single frequency mode (green line with circle marks), in SASE mode based on an optimal $R_{56}$ = 0.3 μm (blue line with square marks), and in SASE mode with a reduced $R_{56}$ = 0.1 μm (red line with diamond marks).

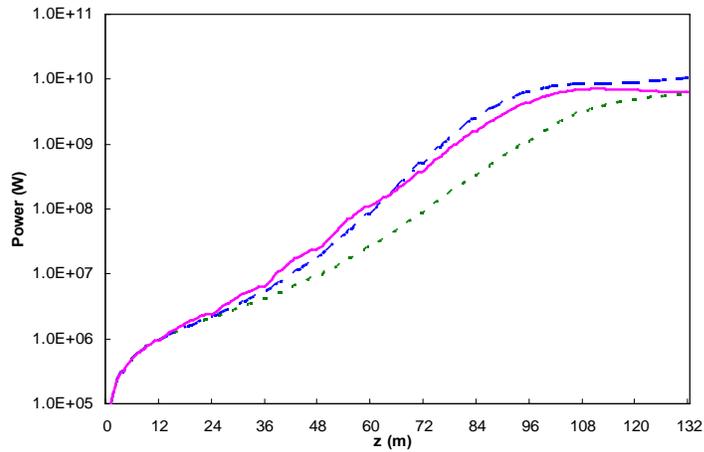

Fig. 7. (color) SASE FEL Power along the undulator at the electron beam emittance of 1.5 μm without any optical klystron (green dotted curve) and with 4 optical klystrons (pink solid curve), and at the emittance of 1.2 μm without any optical klystron (blue dashed curve) . FEL wavelength is 1.5 Å, the undulator parameter K is 3.5, and the initial electron rms energy spread is $5\times10^{-5}$.

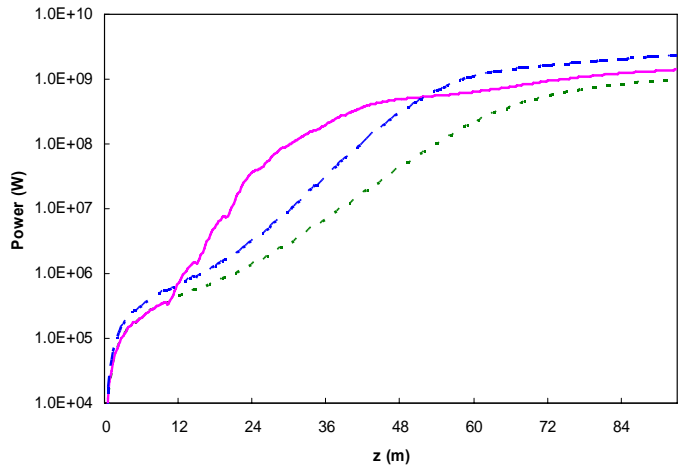

Fig. 8. (color) SASE FEL power along the undulator at a peak current of 2kA without any optical klystron (green dotted curve) and with 4 optical klystrons (pink solid curve), and at peak current of 3kA without any optical klystron (blue dashed curve). The FEL wavelength is 1.5Å and the undulator parameter K=1.3.

[1] Linac Coherent Light Source (LCLS) Conceptual Design Report, SLAC Report No. SLAC-R-593, 2002.

[2] Technical Design Report, DESY TESLA-FEL Report, 2002.

[3] SCSS X-FEL conceptual design report, RIKEN, 2002.

[4] N.A. Vinokurov and A.N. Skrinsky, Preprint of INP **77**-59, Novosibirsk, (1977).

[5] Y.K. Wu, et al., Proceedings of the 2005 Free Electron Laser conference, Stanford, US, 407, (2005).

[6] R. Bonifacio, R. Corsini, and P. Pierini, Nucl. Instrum. Methods Phys. Res. Sec. A **45**, 4091 (1992).

[7] N.A. Vinokurov, Nucl. Instrum. Methods Phys. Res. A **375**, 264 (1996).

[8] K.J. Kim, Nucl. Instrum. Methods Phys. Res. A **407**, 126 (1998).

[9] G.R. Neil and H.P. Freund, Nucl. Instrum. Methods Phys. Res. Sec. A **475**, 381 (2001).

[10] M. Hüning and H. Schlarb, in Proceedings of the 2003 Particle Accelerator Conference, Portland, 2074 (2003).

[11] E.L. Saldin, E.A. Schneidmiller, and M.V. Yurkov, Proceedings of the 2004 Free Electron Laser Conference, Trieste, Italy, 143, (2004).

[12] R. Bonifacio, C. Pellegrini, and L.M. Narducci, Opt. Commun. **50**, 373 (1984).

[13] R. Bonifacio, L. De Salvo, P. Pierini, N. Piovella, and C. Pellegrini, Phys. Rev. Lett. **73**, 70 (1994).

[14] E.L. Saldin, E.A. Schneidmiller, and M.V. Yurkov, Opt. Commun. **148**, 383 (1998).

[15] Z. Huang, D. Dowell, P. Emma, C. Limborg-Deprey, G. Stupakov, and J. Wu, Proceedings of the 2005 Particle Accelerator Conference, Tennessee, 3570, (2005).

[16] M. Borland, Y. Chae, P. Emma, J. Lewellen, V. Bharadwaj, W. Fawley, P. Krejcik, C. Limborg, S. Milton, H.-D. Nuhn, R. Soliday, and M. Woodley, Nucl. Instrum. Methods Phys. Res. A **483**, 268 (2002).

[17] E.L. Saldin, E.A. Schneidmiller, and M.V. Yurkov, Nucl. Instrum. Methods Phys. Res. A **490**, 1 (2002).

[18] S. Heifets, G. Stupakov, and S. Krinsky, Phys. Rev. ST Accel. Beams **5**, 064401 (2002).

[19] Z. Huang and K.-J. Kim, Phys. Rev. ST Accel. Beams **5**, 074401 (2002).

[20] E.L. Saldin, E.A. Schneidmiller, and M.V. Yurkov, DESY Report No. TESLA-FEL-2003-02, 2003.

[21] Z. Huang, M. Borland, P. Emma, J. Wu, C. Limborg, G. Stupakov, and J. Welch, Phys. Rev. ST Accel. Beams **7**, 074401 (2004).

[22] P. Emma, Z. Huang, C. Limborg-Deprey, J. Wu, W. Fawley, M. Zolotorev, and S. Reiche, the Proceedings of 2005 Particle Accelerator Conference, Tennessee, 344, (2005).

[23] E.L. Saldin, E.A. Schneidmiller, and M.V. Yurkov, Nucl. Instrum. Methods Phys. Res. A **381**, 545 (1996).

[24] S. Reiche, Nucl. Instrum. Methods Phys. Res. A **429**, 243 (1999).